\pdfoutput=1
\documentclass[%
 reprint,
 amsmath,amssymb,
 aps,
]{revtex4-1}

\usepackage{graphicx}
\usepackage{dcolumn}
\usepackage{bm}
\usepackage{amsmath}
\usepackage{float}
\usepackage[T1]{fontenc}
\usepackage[utf8]{inputenc}
\usepackage{hyperref}
\hypersetup{
    colorlinks,
    citecolor=blue,
    linkcolor=blue,
    urlcolor=blue
}

\newcommand{\dd}{\mathrm{d}}
\newcommand{\pp}{\varphi}
\newcommand{\mQsq}{\mathcal{Q}^2}
\newcommand{\mQ}{\mathcal{Q}}
\newcommand{\RN}{Reissner–Nordstr\"{o}m }
\newcommand{\Rii}{$\mathcal{R}_2$ }
\newcommand{\Ri}{$\mathcal{R}_1$ }
\newcommand{\Riin}{$\mathcal{R}_2$}
\newcommand{\Rin}{$\mathcal{R}_1$}

\newcommand{\ed}[1]{{#1}}

\begin{document}

\preprint{APS/123-QED}

\title[Reflection-asymmetric wormholes and their double shadows]{Reflection-asymmetric wormholes and their double shadows}

\author{Maciek Wielgus}
 \email{maciek.wielgus@gmail.com}
	\affiliation{Black Hole Initiative at Harvard University, 20 Garden Street, Cambridge, MA 02138, USA}
		\affiliation{Center for Astrophysics  $|$ Harvard \& Smithsonian, 60 Garden Street, Cambridge, MA 02138, USA}
\author{Ji\v{r}\'{i} Hor\'{a}k}
\affiliation{Astronomical Institute, Academy of Sciences, Bo\v{c}ni II 1401, CZ-14131 Prague, Czech Republic}
\author{Frederic H. Vincent}
\affiliation{LESIA, Observatoire de Paris, Universit\'e PSL, CNRS, Sorbonne Universit\'es, UPMC Univ. Paris 06, Univ. de Paris, Sorbonne Paris Cit\'e, 5 place Jules Janssen, 92195 Meudon, France}

\author{Marek Abramowicz}
\affiliation{Nicolaus Copernicus Astronomical Centre, Polish Academy of Sciences, Bartycka 18, 00-716 Warsaw, Poland}
	\affiliation{Research Center for Computational Physics and Data Processing; Institute of Physics, Silesian University in Opava, 
	Czech Republic}
	\affiliation{Department of Physics, G{\"o}teborg University, Sweden }


\date{\today}

\begin{abstract}
We discuss construction and observational properties of wormholes obtained by connecting two Reissner–Nordstr\"{o}m spacetimes with distinct mass and charge parameters. These objects are spherically symmetric, but not reflection-symmetric, as the connected spacetimes differ. The reflection-asymmetric wormholes may reflect a significant fraction of the infalling radiation back to the spacetime of its origin. We interpret this effect in a~simple framework of the effective photon potential. Depending on the model parameters, image of such a wormhole seen by a distant observer (its "shadow") may contain a~photon ring formed on the observer's side, photon ring formed on the other side of the wormhole, or both photon rings. These unique topological features would allow us to firmly distinguish this class of objects from \ed{Kerr} black holes using radioastronomical observations.
 
\end{abstract}

\maketitle

\section{Introduction}
    \label{sec:intro} 
    Traversable wormholes are spacetime tunnels connecting universes or distant parts of the same universe, through which transit of mass and energy is possible. They were proposed and discussed by \citet{Ellis1973} and later by \citet{Morris1988}. A~particularly simple construction, a~symmetric wormhole obtained by surgically grafting two Schwarzschild spacetimes (cut-and-paste procedure), was given by \citet{Visser1989}. A~thin spherical layer of exotic matter (violating the weak energy condition of non-negative energy density), concentrated at the junction between the two connected spacetimes, is required to fulfill the Einstein field equations and to stabilize the wormhole. Similar requirements are common to a~broader class of wormholes consistent with the general relativity \cite{Poisson1995}. \ed{Alternative theories of gravity \cite{Gravanis2007,Harko2013,Moraes2018}, or general relativity in higher number of dimensions \cite{Svitek2018}, may admit wormhole solutions without invoking the exotic stress-energy tensor.}
    
    In recent years, a lot of research has been dedicated to calculating the appearance of wormholes illuminated by the electromagnetic radiation \cite{Nandi2006,Tsukamoto2012,Bambi2013,Nedkova2013,Ohgami2015,Abduj2016,Shaikh2018,Shaikh2019,Amir2019}. This interest has been sparkled, at least in part, by the developments in the radiointerferometry and assemblement of the Event Horizon Telescope (EHT). The EHT is able to resolve the event horizon scale structure for at least two nearby objects, our Galactic Center \cite{Doeleman2008}, and the center of the M87 galaxy \cite{Shep2012,EHT2019p1,Wielgus2020}, with a potential to resolve many more sources in the future  \cite{Haworth2019,Pesce2019}. At this point it becomes possible to observationally distinguish between black holes and certain classes of black hole mimickers \cite{Psaltis2019,EHT2019p5}. Wormholes constitute an important type of the latter, as an example of horizonless spacetimes that may be identical to the black hole spacetime everywhere apart from its most internal part.
    
    \ed{Spacetimes of compact objects may admit unstable spherical null geodesics \cite{Claudel2001,Teo2003}, forming a~photon sphere. The null geodesics approaching these unstable spherical orbits arbitrarily close create a~critical curve on a~distant
    observer's screen \cite{Gralla2019,Johnson2020}. Size and shape of that curve is dictated entirely by the geometry of spacetime, and not the geometry of the source of the radiation. In case of \ed{Kerr} black holes critical curves surrounding a~dark "shadow" region were first rigorously studied by \citet{Bardeen1973} and \citet{Luminet1979}. While this asymptotic feature itself is not observable, in a~realistic astrophysical scenario we expect an image of a~compact object to contain sharp features approximating the critical curves - "photon rings", corresponding to photon trajectories approaching the spherical geodesics \cite{EHT2019p5,Gralla2019,Johnson2020,Vincent2020}. Amount of flux transported along these geodesics is enhanced, as a~consequence of an increased path through the radiation-emitting region surrounding the compact object. Photon rings could be, at least in principle, identified with very high angular resolution radiointerferometric observations \citep{Johnson2020,Vincent2020}. In this paper we study the critical curves, but we occasionally refer to the results as "photon rings" or "shadows", which is somewhat imprecise, but follows a~common convention in the literature. We avoid using the "Einstein ring" term, commonly associated with a~scenario requiring a~very specific geometric arrangement of the system and involving much smaller deflection angles \cite{Schneider1992}.
    }
    
    We discuss critical curves for a~class of wormholes distinguished by the asymmetry between the spacetimes they connect, a~"reflection" asymmetry with respect to the wormhole throat \cite{Montelongo2012,Forghani2018,Wang2020}. As a~representative \ed{model} example, we discuss in more detail wormholes connecting two \RN  spacetimes, following the reflection-symmetric constructions considered by \cite{Visser1989,Eiroa2004,Sharif+Azam2013}. However, in our case \RN spacetimes on both sides of the wormhole are characterized by generally different masses $M_{1,2}$, and charge parameters $\mathcal{Q}_{1,2}$. Because of the spherical symmetry of the spacetimes that we consider, the shadows remain circularly symmetric. Nevertheless, the reflection-asymmetry of the wormhole spacetime has significant consequences for the associated shadow, which may indicate presence of a secondary component, corresponding to the unstable photon sphere from the other side of the wormhole, or even consist exclusively of the component from the other side, that may not match the gravitational signature of the spacetime on the side of the observer. Hence, we define a~class of black hole mimickers that may indicate observational features topologically distinct from that of Kerr black holes, and could potentially be distinguished with the future observations.
    

\section{Effective photon potential of a~wormhole}
\label{sec:Veff}
Let us consider a spherically symmetric spacetime with metric $g_{\mu \nu}$ in spherical coordinates $\{t,r,\theta, \phi\}$,
\begin{equation}
    \dd s^2\!=\!g_{\mu \nu}\!\dd x^\mu \dd x^\nu\!=\!-f \dd t^2 + f^{-1} \dd r^2 + r^2(\dd \theta^2 + \sin^2 \theta \dd \phi^2),
    \label{eq:metric_general}
\end{equation}
where the function $f \equiv f(r)$ will be specified later and we employ the $(-+++)$ signature. For an equatorial null geodesic, it follows from the condition $p^\mu p_\mu = 0$ that
\begin{figure}[t!]
    \centering
    \includegraphics[width=0.99\linewidth]{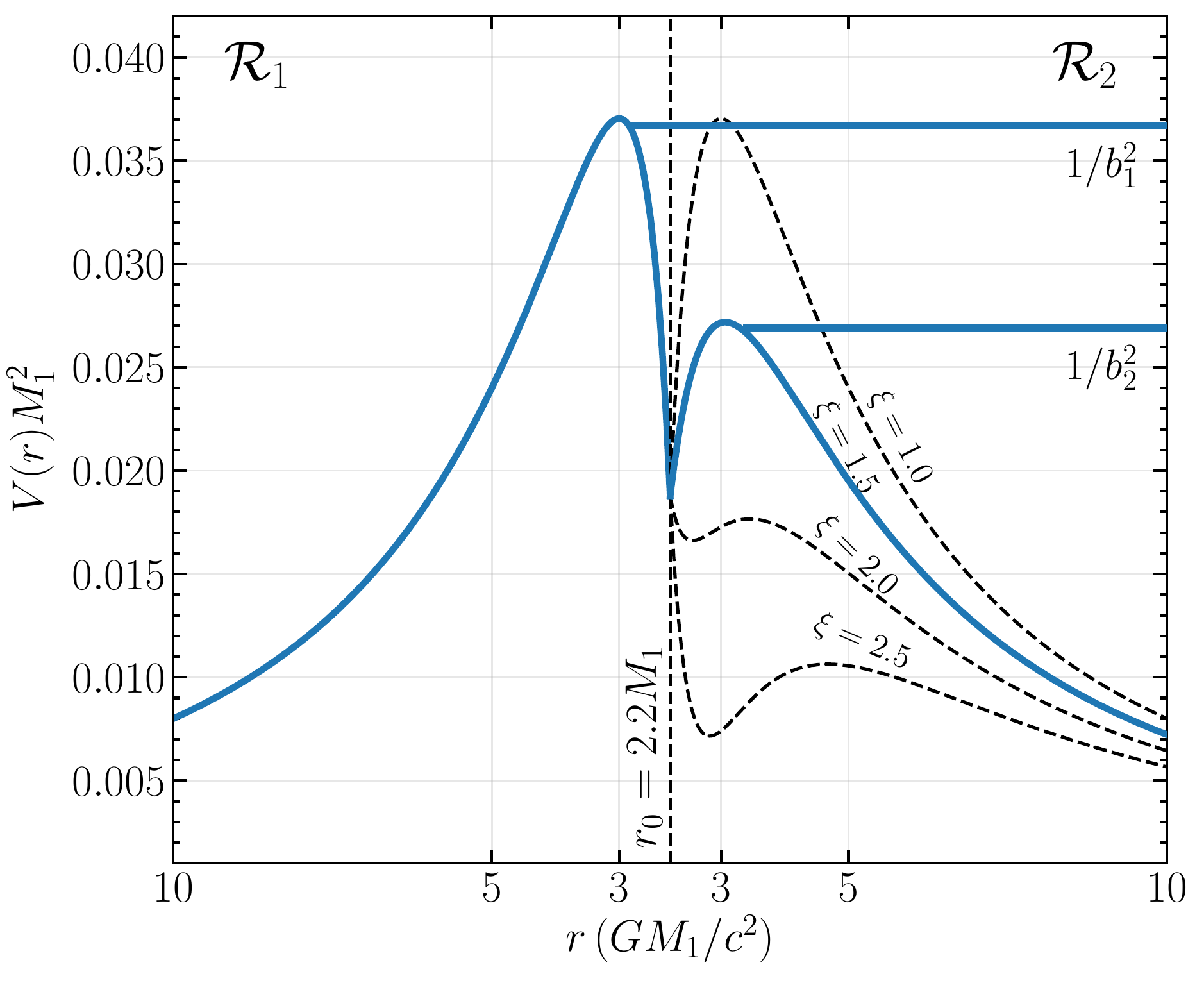}
    \caption{Effective photon potential $V(r)$ for the asymmetric wormhole, obtained by connecting two manifolds $\mathcal{R}_1$ and $\mathcal{R}_2$. $\mathcal{R}_1$ is a~fixed Schwarzschild spacetime, while $\mathcal{R}_2$ is a~Reissner-Nordstr\"{o}m spacetime matched for several values of an asymmetry parameter $\xi = M_2/M_1$. As an example, for $\xi=1.5$, two photons with impact parameters $b_1$ and $b_2$ are shown. The photon with impact parameter $b_2$ is reflected at the effective-potential barrier as in ordinary \RN spacetime before it reaches the throat at $r_0$. On the other hand, the photon corresponding to $b_1$ crosses the potential barrier in $\mathcal{R}_2$, falls into the wormhole, but then it is reflected back to $\mathcal{R}_2$ by the $\mathcal{R}_1$ potential barrier.}
    \label{fig:RNWpotentials}
\end{figure}
\begin{equation}
    \frac{p_t^2}{f} - \frac{p_\phi^2}{r^2} = \frac{\left(p^r\right)^2}{f},
    \label{eq:u-norm}
\end{equation}
where $p^\mu = \dd x^\mu/ \dd s$ is a~photon four-momentum and $s$~is a~properly chosen affine parameter. The components $p_t$ and $p_\phi$ are conserved along the geodesic due to the Killing symmetries of the considered spacetime. Their ratio $b = - p_\phi/p_t$ is the impact parameter of the photon (also referred to as a~specific angular momentum). Eq.~(\ref{eq:u-norm}) can be rearranged in the form
\begin{equation}
   \frac{1}{b^2} - \frac{f}{r^2} = \frac{1}{r^4}\left(\frac{\dd r}{\dd\phi}\right)^2\geq 0.
\label{eq:derive_potential}
\end{equation}
The second term on the left-hand side
\begin{equation}
    V(r) = \frac{f(r)}{r^2}
    \label{eq:effective_potential}
\end{equation}
plays a role of an effective potential - a photon with an impact parameter $b$ can propagate only in the regions where $1/b^2 \geq V(r)$. The turning points correspond to $1/b^2 = V$, hence the radial location of the maximum of the effective photon potential corresponds to the unstable photon orbit and the value of $b$ at the potential maximum is the radius of the observed photon ring. The effective photon potential is thus a~useful tool to diagnose the black hole shadow. The shape of the effective photon potential is also relevant in the context of gravitational wave ringdowns, as discussed, e.g., by \citet{Cardoso2016}, who explored a~symmetric Schwarzschild wormhole case, and more recently by Hor\'{a}k {\it{et al.}} (in prep), who discussed ultra-compact stars.

Figure \ref{fig:RNWpotentials} shows the effective potential of a wormhole connecting two manifolds $\mathcal{R}_1$ and $\mathcal{R}_2$ at a throat located at $r=r_0$. Here $\mathcal{R}_1$ is a Schwarzschild spacetime. We denote $\xi = M_2/M_1$. For $\xi = 1$ we find a thin-shell symmetric wormhole of \citet{Visser1989}. The critical curve is formed by photons approaching the effective potential maximum. As long as the potential barrier is the same on both sides of the throat, the shape of the critical curve will be consistent with that corresponding to a~black hole of mass $M_1$. However, if we construct a~wormhole with an asymmetric effective potential, such as the blue curve $\xi = 1.5$ in Fig.~\ref{fig:RNWpotentials}, the situation will change dramatically. The observers in $\mathcal{R}_2$ should see a~shadow associated with the effective potential maximum in $\mathcal{R}_2$, consistent with the expectations for the $M_2$ black hole, and formed by photons of impact parameter $\approx b_2$. However, they will also see a~shadow feature associated with the photon effective potential maximum in $\mathcal{R}_1$, formed by photons with an impact parameter $\approx b_1$, of an unexpected diameter inconsistent with the expectations for the $M_2$ mass black hole. In Sec. \ref{sec:construction}, we show that such asymmetric effective potentials can be constructed by considering wormholes connecting \RN spacetimes.

\section{Reissner-Nordstr\"om spacetime}
In the case of the \RN spacetime, the function $f(r)$ in Eq.~(\ref{eq:metric_general}) is given by
\begin{equation}
    f(r) = 1 - \frac{2 M}{r} + \frac{\mQsq}{r^2},
\end{equation}
where $M$ and $\mQ$ are the mass and the electric charge parameters, respectively. For $\mQ^2 \leq M^2$, the condition $f=0$ implies presence of two horizons, located at
\begin{equation}
r_{\rm h \pm} = M \pm (M^2 - \mQsq)^{1/2},
\end{equation}
and the metric describes a charged non-rotating black hole. \ed{We refer to the larger one, $r_{\rm h +}$, as an event horizon, while $r_{\rm h -}$ is a~Cauchy horizon.} We will denote the event horizon radius with $r_{\rm h}$, that is $r_{\rm h} \equiv r_{\rm h +}$. For $\mQ^2 > M^2$, \RN metric describes a~spherically-symmetric charged naked singularity.
Photon sphere \ed{(location of the spherical null geodesics)} radius $r_\gamma$ 
follows from the condition $ \dd V / \dd r = 0$, which leads to the quadratic equation
\begin{equation}
    r_\gamma^2 - 3M r_\gamma + 2 \mQsq = 0
\end{equation}
with two roots
\begin{equation}
    r_{\gamma \pm} = \frac{3M \pm (9 M^2 - 8\mQsq)^{1/2}}{2}.
\end{equation}
Larger solution $r_{\gamma +}$ corresponds to a local maximum of $V(r)$, related to the unstable photon orbit.
Note also that the solutions $r_{\gamma\pm}$  exist only for $\mQsq \le \frac{9}{8}M^2$. We will denote the larger root with $r_{ \gamma}$, that is $r_{\gamma} \equiv r_{\gamma +}$.

\begin{figure}[t!]
    \centering
    \includegraphics[width=0.99\linewidth]{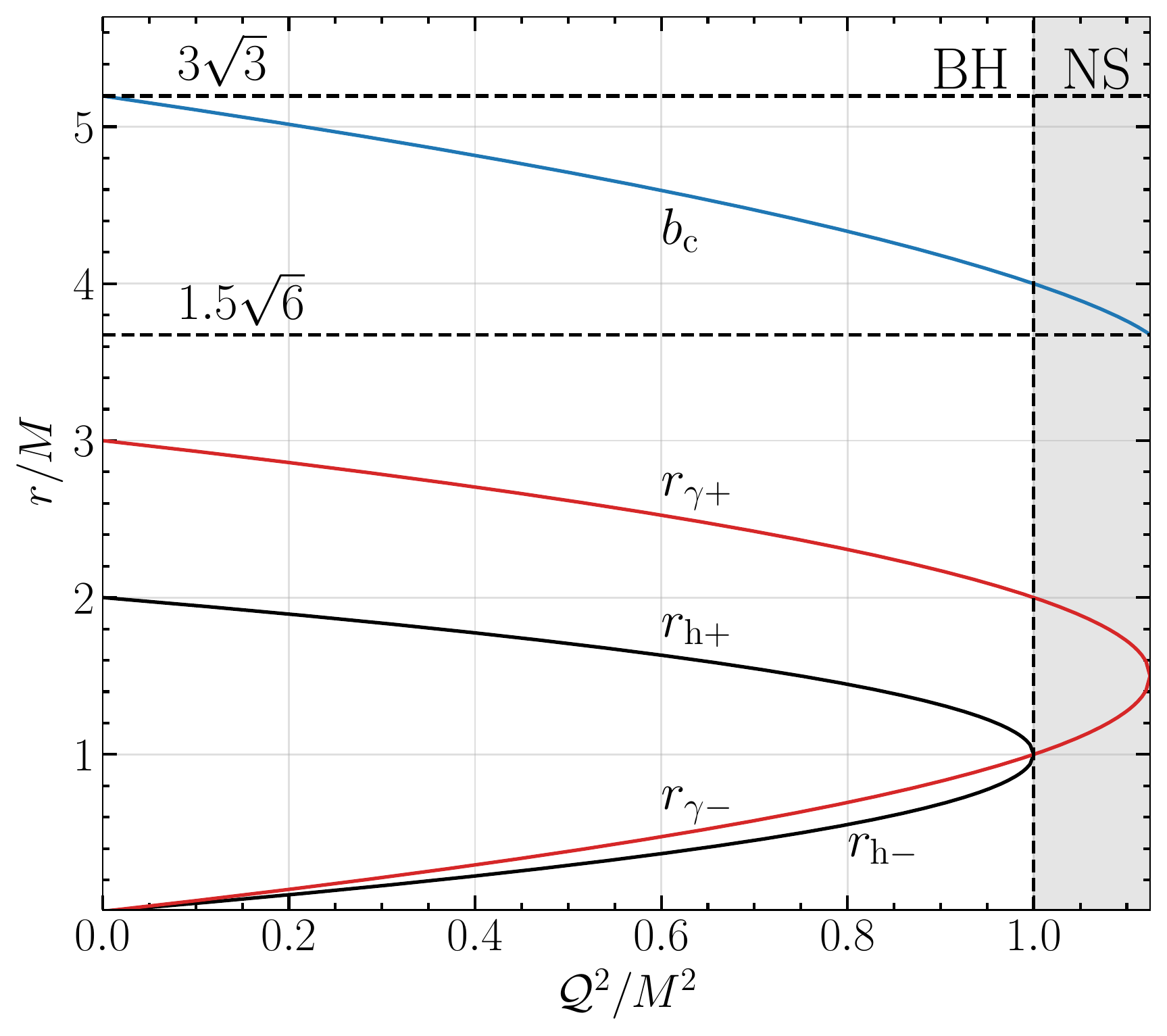}
    \caption{Relevant radii in the \RN spacetime: event horizons $r_{\rm h \pm}$, photon spheres $r_{\rm \gamma \pm}$, and the critical impact parameter $b_{\rm c}$. For $\mQsq \le M^2$ the spacetime corresponds to a charged black hole. The shaded region for $1 < \mQsq \le 9/8 M^2$ corresponds to a naked singularity solution.}
    \label{fig:RNstructure}
\end{figure}

Radius of the critical curve (shadow seen by a distant observer) is given by the critical impact parameter $b_{\rm c}$, corresponding to the maximum of the effective potential,
\begin{equation}
    b_{\rm c} = V_\mathrm{max}^{-1/2} = r_{\gamma} [f(r_{\gamma})]^{-1/2}.
\label{eq:bc}
\end{equation}
Location of the horizons $r_{\rm h \pm}$, photon spheres $r_{\gamma \pm}$, and the value of the critical impact parameter $b_{\rm c}$ as functions of $\mathcal{Q}^2/M^2$ are shown in Fig.~\ref{fig:RNstructure}. All the relevant radii $r_{\rm h}$, $r_{\rm \gamma}$, $b_{\rm c}$ decrease monotonically with $\mathcal{Q}^2$ \cite{Zakharov1994}. Impact parameter decreases by $\sim$ 30\% between $\mathcal{Q}^2/M^2 = 0$ and $\mathcal{Q}^2/M^2 = 9/8$. In certain alternative theories of gravity \cite{Dadhich2000} negative $\mathcal{Q}^2$, reinterpreted as a gravitational "tidal charge", is admitted and yields larger $b_{\rm c}$. However, we limit our discussion to $0 \le \mathcal{Q}^2 \le 9/8 M^2$.

\ed{In a realistic astrophysical context it is unlikely that a~black hole could maintain an electric charge yielding $\mathcal{Q}^2$ comparable to $M^2$ \citep{Wald1974}. However, it remains very unclear, what would constitute a~realistic astrophysical context for an exotic object such as a~traversable wormhole. Regardless of these concerns, a~\RN spacetime and \RN wormholes constitute a~very useful simple model to explore the deviations from the Schwarzschild case when $\sim\!r^{-2}$ term in $f(r)$ is introduced.}

    
\section{Matching Reissner-Nordstr\"om spacetimes}
\label{sec:construction}
It has been first noticed by \citet{Visser1989} that a~traversable wormhole can be formed using a~simple cut-and-paste technique applied to two Schwarzschild spacetimes. The necessary condition of matching the induced metrics at the junction is trivially fulfilled when Schwarzschild spacetimes corresponding to identical masses are matched at the same Boyer-Lidquist coordinate radius.
The resulting wormhole is therefore reflection-symmetric around the throat and so is the effective photon potential, as discussed in Sec.~\ref{sec:Veff}. 
Reflection-asymmetric wormhole solutions formed by stitching two Schwarzschild and \RN spacetimes were presented by \citet{Montelongo2012}. 
Here we consider a simple cut-and-paste construction of an asymmetric \RN wormhole, where spacetime is static, location of the throat is constant in time, and we can explicitly match the full metric on both sides in Boyer-Lindquist coordinates. Because of the $g_{tt}$ continuity, not demanded by a more general construction of reflection-asymmetric wormholes, we conserve the energy $E = - p_t$ of a photon crossing the wormhole throat. 

\begin{figure}[H]
    \centering
    \includegraphics[trim=3.0cm 4.75cm 3.0cm 4.75cm,clip,width=0.99\linewidth]{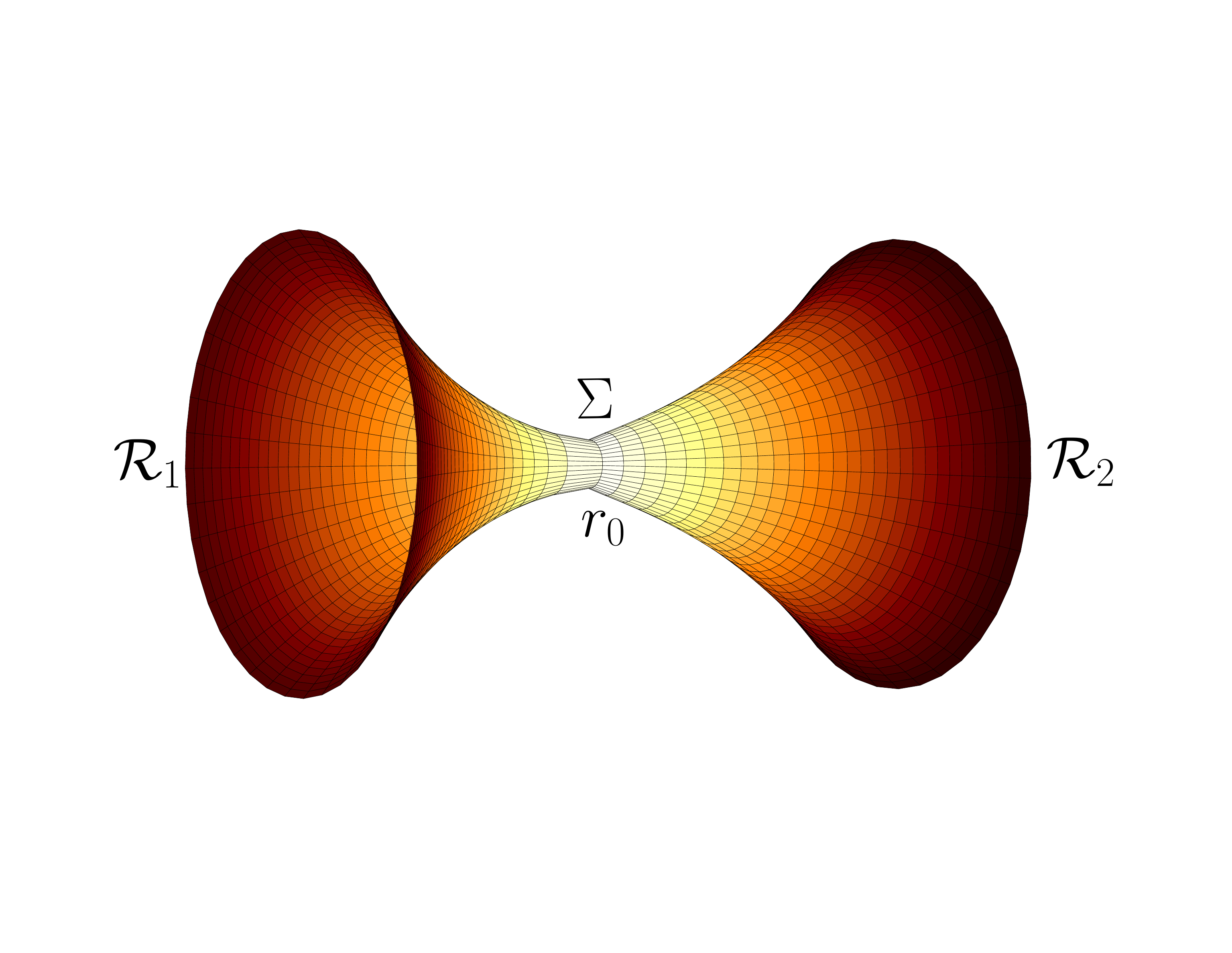}
    \caption{Embedding diagram for a reflection-asymmetric thin-shell traversable wormhole with parameters $M_2 = 1.6M_1$, $r_0 = 2.1 M_1$, $\mathcal{Q}_1^2=0$,$\mathcal{Q}_2^2=0.98 M_2^2$.}
    \label{fig:embedding}
\end{figure}

To outline such a solution, let us consider two manifolds $\mathcal{R}_1$ and $\mathcal{R}_2$ arising from two different \RN spacetimes by cutting-off their interior parts at radii $r_1$ and $r_2$ (respectively), $\mathcal{R}_1 = \{r > r_1 | r_1 > r_{\rm{h},1}\}$, $\mathcal{R}_2 = \{r > r_2 | r_2 > r_{\rm{h},2}\}$, The two manifolds are then glued together by identifying their boundaries, $\partial\mathcal{R}_1\equiv\partial\mathcal{R}_2 \equiv \Sigma$, Fig.~\ref{fig:embedding}. We require the metric coefficients $g_{\mu\nu}$ to remain continuous across the junction, that is $g_{\mu\nu, 1}|_\Sigma$ = $g_{\mu\nu,2}|_\Sigma$. Derivatives of the metric may be discontinuous, reflecting the presence of a~massive and charged thin shell that is the source of gravitational and electromagnetic field \cite[e.g.,][]{Eiroa2004}.
The conditions we impose at the junction are
\begin{align}
    & r_1^2 = r_2^2 \equiv r_0^2 , 
    \label{eq:match_gthth}
    \\
    & f_{1}(r_0 ; M_{1}, \mathcal{Q}_{1}) = f_{2}(r_0 ; M_{2}, \mathcal{Q}_{2}),
    \label{eq:match_gtt}
\end{align}
where
\begin{equation}
    f_{1,2}(r ; M_{1,2}, \mathcal{Q}_{1,2})  = 1 - \frac{2 M_{1,2}}{r} + \frac{\mathcal{Q}_{1,2}^2}{r^2}.
\end{equation}
Condition (\ref{eq:match_gthth}) assures continuity of $g_{\phi\phi}$ and $g_{\theta\theta}$ components, while the later one (\ref{eq:match_gtt}) is required by the continuity of $g_{tt}$ and $g_{rr}$. Note that under these assumptions matching Schwarzschild spacetimes with $M_1 \neq M_2$ is not possible, as $f_1(r_0,M_1,0)=f_2(r_0,M_2,0)$ implies $M_1 = M_2$.
Introducing the asymmetry parameter 
\begin{equation}
    \xi = M_2/M_1
\end{equation} we can now consider $r_0$, $M_1$, $\xi$, and $\mQ_1$ to be fixed model parameters, and use Eq.~(\ref{eq:match_gtt}) to solve for the charge parameter $\mathcal{Q}_2^2$,
\begin{equation}
    \mathcal{Q}_2^2 = 2 r_0 M_1 (\xi - 1) + \mathcal{Q}_1^2.
    \label{eq:Q2}
\end{equation}
Hence, we can fulfill conditions (\ref{eq:match_gthth}) and (\ref{eq:match_gtt}) for $M_1 \neq M_2$ if we consider \RN wormholes. Our construction constitutes a subset of the solutions considered by \citet{Montelongo2012} and \citet{Forghani2018}, who also studied their stability and related properties of the exotic matter concentrated at the throat. Instead, in the current paper we are interested in the appearance of the reflection-asymmetric wormholes to a distant observer. 

\section{Wormhole appearance}

\begin{figure}[t!]
    \centering
    \includegraphics[width=0.99\linewidth]{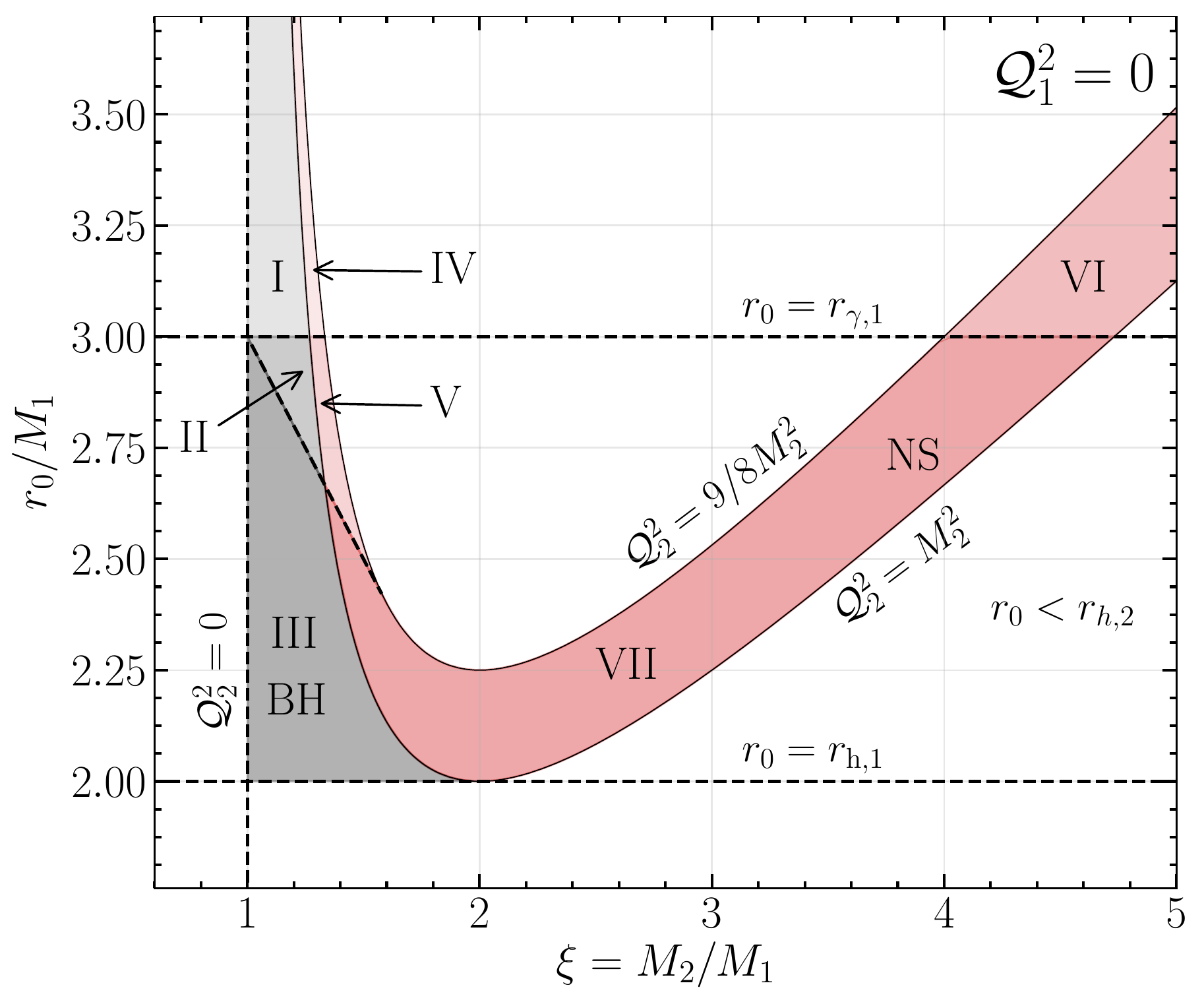}
    \caption{Parameter space for $\mathcal{Q}_1 = 0 $ with varying $r_0/M_1$ and $\xi$. $\mathcal{R}_2$ can be a subset of a \RN charged black hole spacetime (gray-shaded regions I-III) or a subset of a \RN naked singularity spacetime (red-shaded regions IV-VII). Roman numerals denote presence of the photon sphere in $\mathcal{R}_{1,2}$ or lack thereof, see the text. }
    \label{fig:RNWtopoologies}
\end{figure}

We investigate in detail the parameter space in case of $\mathcal{Q}_1^2 \equiv 0$, so when $\mathcal{R}_1$ is a subset of a Schwarzschild spacetime. In Fig.~\ref{fig:RNWtopoologies} we see that this slice of the full parameter space is already very rich in terms of the wormhole shadow topologies. Depending on a~combination of $\xi = M_2/M_1$ and $r_0/M_1$, manifold $\mathcal{R}_2$ can be a subset of a charged black hole (shaded gray) or a~naked singularity (shaded red) spacetime. We first classify the wormhole solutions in terms of presence of the photon orbit in $\mathcal{R}_{1,2}$, so whether $r_0 < r_{\gamma}$. As a result, possible cases indicated in Fig.~\ref{fig:RNWtopoologies} are:
\begin{enumerate}
    \item regions I and IV: $r_0 > r_{\gamma,1}$ and $r_0 > r_{\gamma,2}$, no photon sphere neither in $\mathcal{R}_1$ nor in $\mathcal{R}_2$,
     \item regions II and V: $r_0 < r_{\gamma,1}$ and $r_0 > r_{\gamma,2}$, photon sphere only in $\mathcal{R}_1$,
    \item regions III and VII: $r_0 < r_{\gamma,1}$ and $r_0 < r_{\gamma,2}$, photon spheres in both $\mathcal{R}_1$ and $\mathcal{R}_2$,
     \item region VI: $r_0 > r_{\gamma,1}$ and $r_0 < r_{\gamma,2}$, photon sphere only in $\mathcal{R}_2$.
\end{enumerate}
Presence of an unstable photon orbit on the opposite side of the wormhole is not a~sufficient condition for a distant observer to see the corresponding critical curve - the photons still need to cross the effective photon potential barrier on the observer's side. As an example, in Fig.~\ref{fig:RNWpotentials} a~distant observer in $\mathcal{R}_2$ sees critical curves associated with effective potential maxima in both $\mathcal{R}_1$ and $\mathcal{R}_2$, but the observer in $\mathcal{R}_1$ only sees critical curve associated with the effective potential maximum in $\mathcal{R}_1$. A simple condition for the observer in $\mathcal{R}_{i}$ to observe the critical curve from the other side of the wormhole is therefore given by
\begin{equation}
    \max_{\mathcal{R}_{i}} V_{\rm} (r) < V (r_{\gamma, j}) \ \text{and} \ r_0 < r_{\gamma, j},
\label{eq:see_other_side}
\end{equation}
for $i \neq j$. Because we assume $\mathcal{Q}^2 \ge 0$, it follows from Eq.~(\ref{eq:Q2}) that if $\mathcal{Q}_1 = 0$ then $\xi \ge 1$. In other words, we can not match Schwarzshild spacetime with mass $M_1$ with a \RN spacetime of lower mass $M_2$ within the framework described in Sec.~\ref{sec:construction}. Evaluating numerically condition (\ref{eq:see_other_side}) for the parameter space shown in Fig.~\ref{fig:RNWtopoologies}, we find that under our assumptions the observer in $\mathcal{R}_1$ is never able to see the critical curve related to the photon sphere in $\mathcal{R}_2$. Hence, such an observer may only see the Schwarzschild spacetime critical curve, just as if the observed compact object was a~nonrotating black hole. On the other hand, an observer in $\mathcal{R}_2$ can see the critical curve from $\mathcal{R}_1$ in all cases, as long as $r_0 < r_{\gamma,1} = 3 M_1$. 

The maximum of the effective potential occurs also for spacetime parameters from region I, at the throat of the wormhole. One may argue, that the throat at $r=r_0$ may also correspond to an unstable photon orbit, as it is in the case considered by \citet{Shaikh2019}. However, in the vicinity of the throat, the radial derivative of the effective potential remains finite, having discontinuity across the throat. As a result, the null geodesics do not wind up around $r=r_0$ as in the case of an ordinary unstable photon orbit, where the radial derivatives approach zero. Rather, they are suddenly "reflected" to the other spacetime, creating a discontinuity in the observed image.

\begin{figure}[H]
    \centering
     \includegraphics[width=0.99\linewidth]{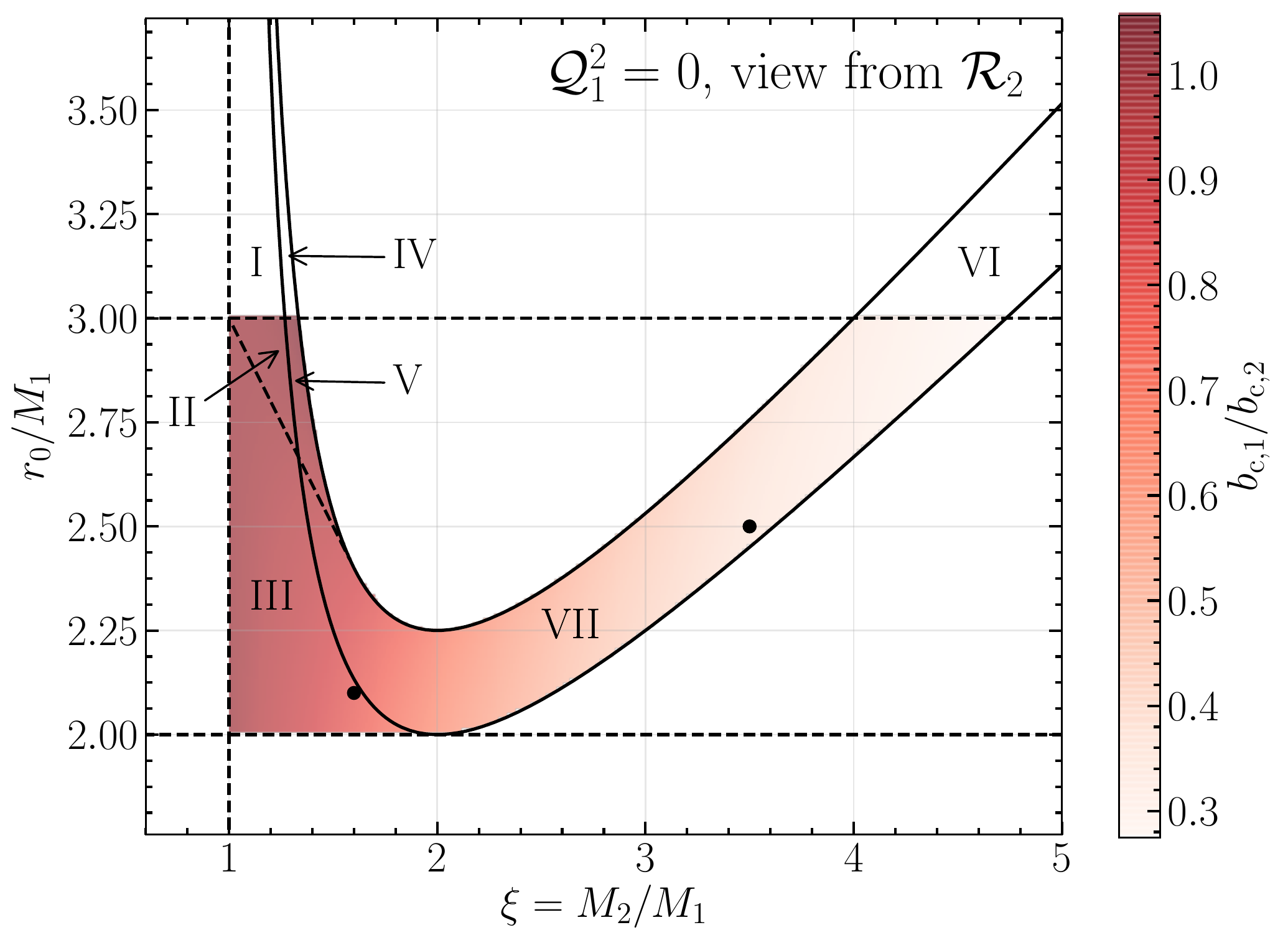}
    \caption{Shaded region corresponds to the part of the parameter space, for which a distant observer in \Rii sees a shadow associated with the photon sphere in \Ri through the wormhole. The color codes ratio of the \Ri shadow radius $b_{\rm c,1}$ with respect to $b_{\rm c,2}$, the expected radius of a~\RN shadow of $\mathcal{R}_2$. Two black dots indicate parameters of the examples considered in Fig.~\ref{fig:trajectories}.}
    \label{fig:alien_shadow}
\end{figure}

\begin{figure*}[t!]
    \centering
    \includegraphics[width=0.99\linewidth]{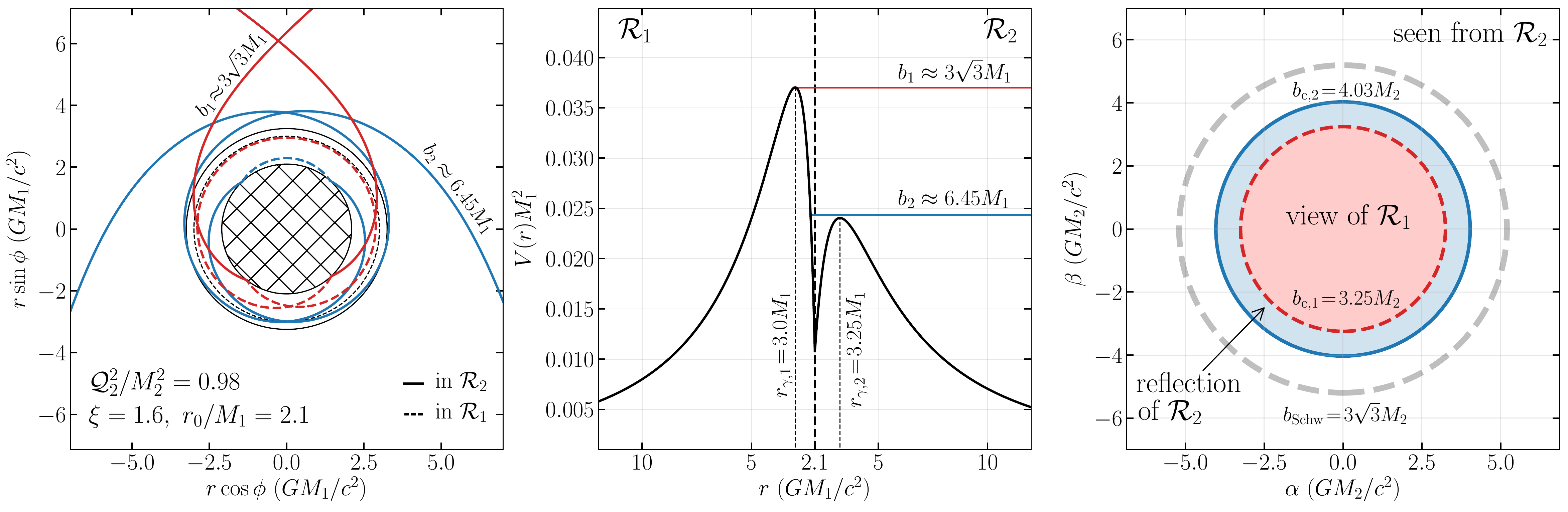}
     \includegraphics[width=0.99\linewidth]{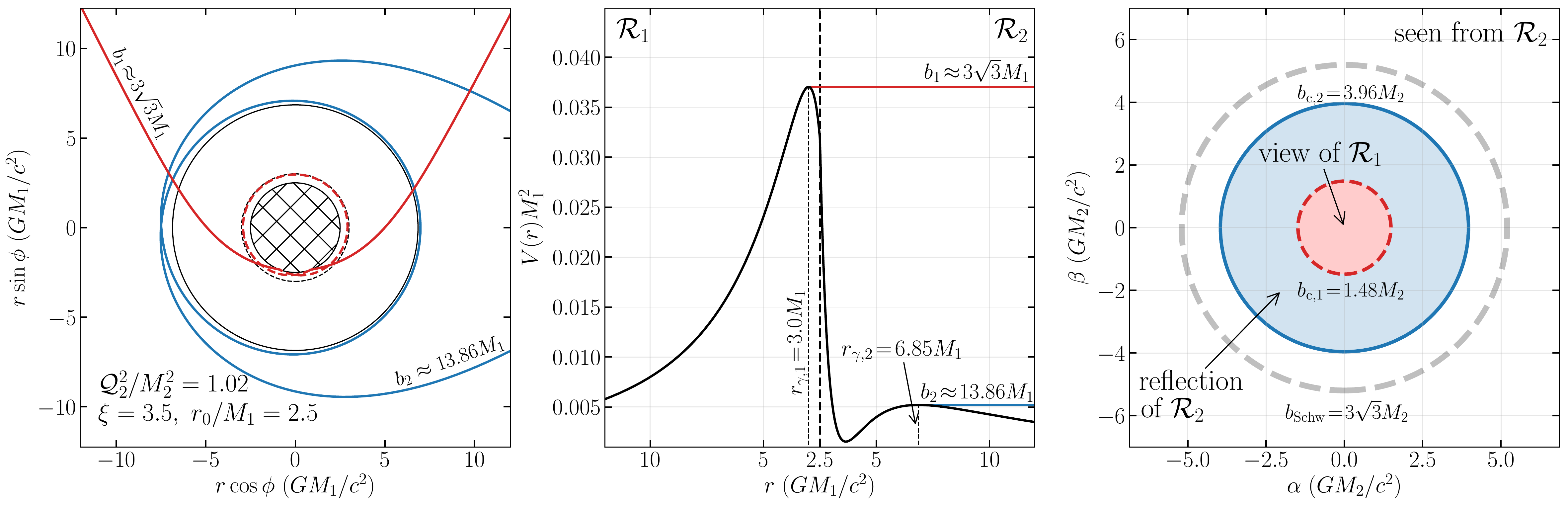}
    \caption{
    \emph{Top row:} Properties of a wormhole connecting a Schwarzschild spacetime \Ri with mass $M_1$ and a \RN spacetime \Rii with mass $1.6 M_1$ and charge $\mathcal{Q}_2^2 = 0.98M_2^2$. Left) trajectories of photons with the impact parameters close to critical values for \Ri and \Riin. Hatched region corresponds to $r < r_0$ and is not a part of the investigated spacetime. Dashed lines indicate trajetories on the other side of the wormhole, that is in $\mathcal{R}_1$. Middle) corresponding reflection-asymmetric effective photon potential. Locations of the unstable photon spheres (maxima of the effective potentials) are indicated. Right) an appearance of the wormhole for a distant observer. Two rings, corresponding to critical curves in \Ri ($b_{\rm c, 1}$) and in \Rii ($b_{\rm c, 2}$), are visible. A~region in which photons visit \Ri and are reflected back to \Rii is shaded in blue. Celestial coordinates $(\alpha, \beta)$ are measured in $G M_2 /c^2$. \emph{Bottom row:} same, but for the \Rii spacetime parameters $M_2 = 3.5 M_1$ and $\mathcal{Q}_2^2 = 1.02 M_2^2$.
    }
    \label{fig:trajectories}
\end{figure*}

In Fig.~\ref{fig:alien_shadow} we evaluate the ratio between the size of the shadow originating in $\mathcal{R}_1$ observed from $\mathcal{R}_2$ and the expected shadow in $\mathcal{R}_2$, that is $b_{\rm c,1}/b_{\rm c,2}$. What this means is that even if $r_0 > r_{\gamma,2}$, we use $b_{\rm c,2}$ computed with Eq.~(\ref{eq:bc}), since a distant observer would not know about the throat location $r_0$ and would reasonably expect to see the shadow of a \RN object. The shadow seen through the wormhole may be as much as three times smaller than the expected one. In regions III and VII these two shadows would appear simultaneously. Two such examples are shown in Fig.~\ref{fig:trajectories}. In the regions II and V the \Rii observer would only see the \Ri shadow as $r_0 > r_{\gamma,2}$, nevertheless for the considered wormhole model its size would be quite close to the \RN \Rii expectations. In region VI only the ordinary shadow of \Rii would be seen.

We find trajectories of photons in wormhole spacetimes constructed in Sec.~\ref{sec:construction} by numerically integrating the null geodesic equations of motion. At the junction $r_0$ we use the fact that the metric is continuous and $p_t$ and $p_\phi$ are conserved, while $p_\theta$ remains 0, as we consider an equatorial motion in a spherically symmetric spacetime. Then $p_r$ only requires the sign reversal from ingoing to outgoing. Examples of photon trajectories are shown in Fig.~\ref{fig:trajectories}. 

In the first row of Fig.~\ref{fig:trajectories}, a spacetime from region III of the parameter space shown in Fig.~\ref{fig:RNWtopoologies}, is considered. The embedding diagram for this particular wormhole was shown in Fig.~\ref{fig:embedding}. We are particularly interested in trajectories approaching the unstable photon sphere on each side of the wormhole. Trajectory $b_1$ corresponds to a~photon emitted in $\mathcal{R}_2$ with an impact parameter slightly larger (so $1/b_1^2$ slightly smaller) than the critical value in \Ri of $ b_{c,1} = 3\sqrt{3} M_1$. The photon falls into a~wormhole from \Rii and crosses the throat. It then loops around the unstable photon sphere in \Ri (top left panel) but ultimately is reflected back into the \Rii by the \Ri effective potential barrier (top middle panel). Photon $b_2$ corresponds to the impact parameter slightly smaller than the critical value in \Rii (or $1/b_2^2$ slightly larger). Therefore it loops around the photon sphere in \Rii (top left panel), but ultimately falls into the wormhole, only to be quickly reflected back to \Rii by the \Ri potential barrier (top middle panel). Top right panel of Fig.~\ref{fig:trajectories} outlines the appearance of the wormhole shadow seen by a~distant observer. Dashed gray line shows the Schwarzschild critical curve for the mass $M_2$. Continuous blue line $b_{\rm c,2}$ is the critical curve for the \Rii \RN spacetime with mass $M_2$ and charge parameter $\mathcal{Q}^2_2 = 0.98M_2^2$ (from Eq.~\ref{eq:Q2}). Dashed red line is the \ed{critical curve} from \Rin, seen through the wormhole, corresponding to that of a~Schwarzschild black hole of mass $M_1$, with radius $b_{\rm c,1}$. Inside this circle there is a~region where a view of the \Ri spacetime is seen through the wormhole (shaded red). Between the two shadow features, a~reflection of the \Rii (shaded blue), formed by the photons that visited \Ri but were reflected back into \Rii by the potential barrier, is seen. A~similar scenario, but with a~wormhole spacetime from the region V of the parameter space is investigated in Fig.~\ref{fig:trajectories}, bottom row. In this case the two \ed{photon rings} are of a~very different size, $b_{\rm c,1} / b_{\rm c,2} = 0.37$. 

    
\section{Discussion}
\label{sec:discussion}

We have presented results characterizing the impact of reflection-asymmetry of the effective photon potential on the appearance of a~wormhole to a~distant observer. As an instructive example we considered a~family of thin-shell, traversable, reflection-asymmetric wormholes, obtained by surgically grafting two \RN spacetimes with a cut-and-paste procedure \cite{Visser1989,Montelongo2012}.

We notice interesting features in the shadow (critical curve related to photon geodesics approaching the unstable photon sphere, as systematically defined and discussed by, e.g., \cite{Gralla2019} and \cite{Johnson2020}) of a~reflection-asymmetric wormhole. Apart from variation of the shadow diameter with respect to the expectations, for certain model parameters, observers on one side of the wormhole may be able to see both the shadow corresponding to the photon sphere on their side, and the shadow corresponding to the photon sphere from the other side of the wormhole. While several authors considered wormhole shadows before \cite[e.g.,][]{Nedkova2013,Ohgami2015,Abduj2016,Shaikh2018} a~critical curve consisting of a~double circle is a~rather uncommon feature in the literature. Nevertheless, similar shadows were recently reported by \citet{Shaikh2019}, who considered reflection-symmetric traversable wormholes with a~secondary maximum of the photon effective potential located at the throat. \citet{Wang2020} discussed shadows of asymmetric Schwarzschild wormholes, for which photon energy $E  =- p_t$ is not conserved when a~photon crosses the throat.

\begin{figure}[t!]
    \centering
     \includegraphics[width=0.99\linewidth]{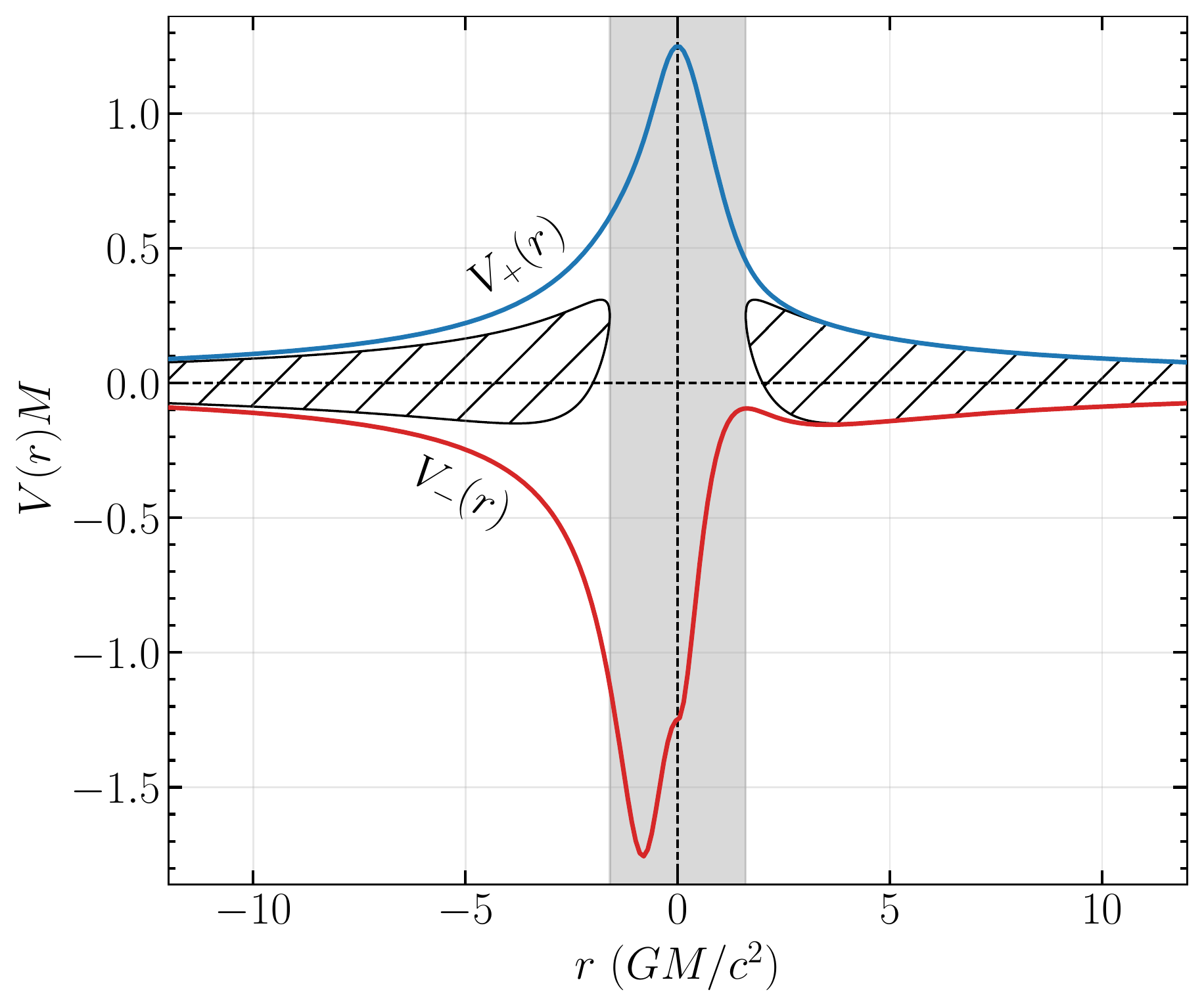}
    \caption{Asymmetric equatorial effective photon potential of a~Lamy wormhole for the parameters considered by \cite{Vincent2020}: dimensionless spin $a_*= 0.8$, charge $b = M$. \ed{There are two extrema of the effective potential $V_{-}$ associated with unstable spherical photon orbits and affecting retrograde photons, one at about 3.5$M$ and other at about -0.8$M$. Potential $V_{+}$ has a single maximum at 0$M$. For comparison, hatched region represents the forbidden region between $V_{-}$ and $V_{+}$ for a~Kerr spacetime. Shaded region indicates the interior of the Kerr event horizon, $|r| < 1.6M$. }
    }
    \label{fig:Lamy}
\end{figure}

\citet{Vincent2020} presented ray-traced images of \ed{several types of black hole mimickers surrounded by an accretion disk, including a~Lamy spinning wormhole \cite{Lamy2018}. In Fig.~10 and Fig.~B.1 of \cite{Vincent2020} images similar to the ones described in this paper (that is, the shadow appearing as multiple circular features) can be seen. The Lamy metric is identical to that of Kerr, with the
only difference that the mass of the object $M$ is replaced
by a~function of the radial coordinate $M(r)$. For large $|r|$ Lamy metric approaches Kerr. In either case two effective photon potentials for the equatorial plane can be defined
\begin{equation}
V_\pm \equiv V_\pm(r) = \frac{-g_{t\pp} \pm \sqrt{g_{t\pp}^2 - g_{tt}g_{\pp\pp}}}{g_{\pp\pp}} \ .
\end{equation}
 In Fig.~\ref{fig:Lamy} we show the $V_\pm(r)$ functions for the parameter values considered in \citep{Vincent2020}. The motion of photons in the equatorial plane is restricted to the region above $V_+$ or below $V_{-}$. The asymmetry of the Lamy potentials $V_\pm(r)$ with respect to $r=0$ leads to multiple critical curves on sky. We however note that this analysis is only valid in the equatorial plane, and that equatorial
geodesics only matter for an exactly edge-on view. A~more detailed treatment is necessary in order to fully understand the
non-edge-on images, such as those presented in \cite{Vincent2020}. Nevertheless, we can conclude that the Lamy wormhole is another example of a~reflection-asymmetric wormhole spacetime, indicating multiple critical curves.}

\ed{Several authors discussed observational constraints on existence of wormholes derived from variety of phenomena \citep{Torres1998,Takahashi2013,Zhou2016,Dai2019,DeFalco2020,Simonetti2020}. Observing geometry of the shadows of compact objects is a~particularly promising avenue.}
Features such as a~presence of multiple \ed{critical curves}, topologically different from the "classic" \ed{Kerr} black hole shadow \cite{Bardeen1973,Luminet1979}, could potentially constitute a~much more powerful discriminant of black hole mimickers than a~moderate difference in \ed{the critical curve} size and circularity. This is particularly important in view of significant uncertainties related to distance and mass of sources that could be potentially resolved by future extremely long baseline radiointerferometry observations \cite{Haworth2019}, perhaps with a~single exception of our Galactic Center. \ed{However, while detection of multiple bright rings on sky would indicate that the observed compact object is not a~Kerr black hole, it would not necessarily imply the non-existence of an event horizon. A~counterexample is given by the black holes with scalar hair, as studied by \citep{Herdeiro2014,Cunha2016,Vincent2016}.}

\ed{Another property of a~traversable wormhole image is a~presence of a~central region directly illuminated by the low angular momentum photons from the other side of the wormhole. Detecting such a~feature would also potentially constitute a~robust discriminant of some black hole mimickers \citep{EHT2019p5}.}

\ed{Apart from these properties,} images of reflection-asymmetric wormholes would contain a~region in which photons emitted on one side of the wormhole visit the other side and are reflected back to the side of their origin (blue-shaded region in the right column of Fig.~\ref{fig:trajectories}). If such a region would ever be observed, its presence could potentially allow for probing the geometry on the other side of a wormhole through investigating delays between the directly observed and reflected events. Such a~special region is exclusively present in the images of the reflection-asymmetric wormholes. \ed{ Fundamentally, it is not even necessary to resolve the source in order to investigate this property - it could manifest as a~delayed correlated component in the compact object's lightcurve. Search for such a~feature could already be performed with the existing lightcurve databases.}

    
\begin{acknowledgments}
 We thank Danial Forghani, Eric Gourgoulhon, Otakar Sv\'{i}tek, and Ronaldo Vieira for useful comments. MW wishes to thank the Astronomical Institute in Prague for hospitality. This work was supported in part by the \textsc{inter-excellence} project No. LTI17018 aimed to strengthen international collaboration of Czech scientific institutions, and the Black Hole Initiative at Harvard University, which is funded by grants from the John Templeton Foundation and the Gordon and Betty Moore Foundation to Harvard University.
\end{acknowledgments}


\bibliography{wormholes}

\end{document}